\documentclass[ApJ]{aastex631}
\usepackage{graphicx,amsmath}

\newcommand{\lapprox} {\, \lower3pt\hbox{$\sim$}\llap{\raise2pt\hbox{$<$}}\,}
\newcommand{\gapprox} {\, \lower3pt\hbox{$\sim$}\llap{\raise2pt\hbox{$>$}}\,}

 \renewcommand{\vec}[1]{\mathbf{#1} }

\begin{document}

\title{Advection-nonlinear-diffusion model of flare accelerated electron
transport in Type III solar radio bursts}

	\author[0000-0002-8078-0902]{Eduard P. Kontar}
	\affil{School of Physics \& Astronomy, University of Glasgow, G12 8QQ, Glasgow, UK}
	\author[0009-0001-7368-0938]{Francesco Azzollini}
  \affil{School of Physics \& Astronomy, University of Glasgow, G12 8QQ, Glasgow, UK}
  \author[0000-0003-3993-2991]{Olena Lyubchyk}
	\affil{Main Astronomical Observatory, Kyiv, Ukraine}
 \affil{Bogomolets National Medical University, Kyiv, Ukraine}


\begin{abstract}
Electrons accelerated by solar flares and observed as type III solar radio
bursts are not only a crucial diagnostic tool for understanding electron
transport in the inner heliosphere but also a possible early indication of
potentially hazardous space weather events. The electron beams travelling in the
solar corona and heliosphere along magnetic field lines generate Langmuir waves
and quasilinearly relax towards a plateau in velocity space.
The relaxation of the electron beam over the short distance in contrast
to large beam-travel distances observed is often referred to as Sturrok's dilemma.
Here, we develop a new electron transport model with quasilinear distance/time
self-consistently changing in space and time. The model results
in a nonlinear advection-diffusion equation for the electron beam density
with nonlinear diffusion term that inversely proportional to the beam density.
The solution predicts slow super-diffusive (ballistic) spatial
expansion of a fast propagating electron beam.
The model also provides the evolution of the spectral energy density
of Langmuir waves,
which determines brightness temperature of plasma radiation in solar bursts.
The model solution is consistent with the results of numerical simulation
using kinetic equations and can explain some characteristics
of type III solar radio bursts.
\end{abstract}

\section{Introduction}\label{sec:introduction}

The signatures of accelerated electrons in solar flares are observed over a wide
range of frequencies from low radio frequencies in the interplanetary space to
gamma-ray range at the Sun.  Hard X-ray and radio observations provide the
most direct signatures of electron acceleration and propagation
in the solar atmosphere and in the interplanetary space
\citep[see, e.g.][as reviews]{1985SoPh..100..537L,2011SSRv..159..107H,2017LRSP...14....2B}.
Since the early radio X-ray and in-situ electron observations \citep{1974SSRv...16..189L}
a close association between X-rays and type III solar radio bursts has been
noted
\citep[e.g.][]{1985SoPh..100..537L,2007ApJ...663L.109K,2014A&A...567A..85R}
suggesting that the cloud of non-thermal beam electrons travels
from the solar flare site into the interplanetary space.
These energetic electrons and associated type III emission could be used for
the forecasting of radiation hazards from solar energetic ion events
\citep{2007SpWea...5.5001P}. However, the quantitative description of electron
transport responsible for type III bursts is a long-standing challenge. The
solar flare electrons propagating along open magnetic field lines are believed
to be responsible for type III solar radio bursts via generation of Langmuir waves
and subsequent conversion of these Langmuir waves into escaping radio
emission \citep{1958SvA.....2..653G}.
Because the faster electrons overtake the slower ones, the conditions for
beam-plasma instability quickly appear leading to a plateau in velocity space
and Langmuir waves generation.
The characteristic time of kinetic beam-plasma
instability (quasilinear relaxation) is normally short $\tau_\text{q}\approx
n_p/(n_b\omega_{pe})$, where $\omega_{pe}^2=4\pi e^2n_p/m$ is the electron
plasma frequency, $n_b, n_p$ are the electron densities of beam and plasma
respectively \citep{1961SvPhU...4..332V}. The mean free path for $30 $~keV
electrons is $\lambda_\text{q}=v\tau_\text{q}\sim 100$~km  ($\tau_\text{q}\sim
10^{-3}$~s. ) for typical type III beam parameters $n_b/n_p\sim 10^{-5}$,
$\omega_{pe}/2\pi= 100$~MHz, $v\sim 10^{10}$~cm/s in the solar corona. Fast
quasilinear relaxation produces beam deceleration \citep[][see also
\citep{1973plas.book.....K,1990SoPh..130..201M,1997SSRv...81..143K,2012SSRv..173..459Y,2015PhPl...22k3109T,2021FrASS...7..116A,2023ApJ...949...24K}]{1964NASSP..50..357S},
a problem that has become known as Sturrock's dilemma. Indeed, spacecraft
observations show solar flare energetic electrons are accompanied by type III
solar radio bursts \citep{1970SoPh...12..266L,1970SoPh...15..222F}. There are
two broad approaches to resolve the dilemma: one avenue invokes modification to
quasilinear Langmuir wave generation (e.g. \citet{1974ApJ...190..175P,
1976ApJ...209..912B,2019JGRA..124...68S} attributed the solution to Sturrock’s
dilemma to nonlinear effects introduced by the oscillating two-stream
instability, stabilization by plasma density inhomogeneities
\citep[e.g.][]{1982PhFl...25.1062G,1985SoPh...96..181M},
cyclic Langmuir collapse \citep{2017PNAS..114.1502C}),
while the second avenue
highlights that electron beam is spatially nonuniform,
so Langmuir waves would preferentially be generated at the front of the beam and absorbed at the back
\citep[e.g.][]{1970SvA....14...47Z,1972SoPh...24..444Z,1995PlPhR..21...89M,2000NewA....5...35M}.
Numerical solutions of kinetic equations
\citep{1976SoPh...46..323T,1977SoPh...55..211M,1982SoPh...81..173G,1982SoPh...78..141T,2001A&A...375..629K,2009ApJ...707L..45H,2008JGRA..113.6104L,2013SoPh..285..217R,2014A&A...572A.111R}
broadly support the latter: although quasilinear relaxation flattens the
electron distribution, spatial inhomogeneity of the electron beam allows
electrons to propagate large distances. While the time-consuming numerical
simulations provide important insights, analytic theory is essential to relate
observable properties of type III bursts and electron beam properties.

One major simplification for the challenge is to utilise the smallness of
quasilinear time in comparison to the characteristic
time of the beam and to seek the hydrodynamic description at the timescales
larger than the quasilinear relaxation
\citep[e.g.][]{1970JETP...31..396R,1995PlPhR..21...89M}.
This is a good assumption due to the smallness of the characteristic time
of beam-plasma interaction (the quasilinear time, $\tau_\text{q}$) compared to the
characteristic time of the beam $t$, $\tau_\text{q} \ll t$.
However, due to finite size of the
electron beam, the quasilinear time being inversely proportional to the beam density should change from small (high beam density near beam center) to large (small beam density away from the beam center) values. Furthermore,
the parameter $\tau_\text{q}/t $ becomes a function of space and time through dependency on the beam density.

In the present work we address this theoretical challenge noting that the
quasilinear time has to be explicitly treated as a function of time and space
depending on the electron number density of the electron beam. Thus,
the relaxation process is not going to be taking place at the same rate in all
points of space. This non-linear description of fast diffusion naturally leads
to a more realistic analytical solutions, comparable to the results of the
numerical simulations shown in the literature.

\section{Kinetic description of electrons and Langmuir waves}\label{sec:kinetic
description}

Quasilinear theory describes the propagation of electrons along magnetic field
lines in a weakly magnetized plasma, and the resonant interaction of the
electrons with Langmuir waves e.g. $\omega_{\mathrm{pe}}=kv$, where
$\omega_{\mathrm{pe}}$ is the local plasma frequency, $k$ is the wave number and
$v$ is the velocity. The quasilinear equations
\citep{1963JETP...16..682V,1964AnPhy..28..478D}, provide kinetic description for
electrons and Langmuir waves in type III solar radio bursts. As the electrons
follow magnetic field lines, the reduced field-aligned electron distribution
function $f(v,x,t)=\int f(\vec{v}) d\vec{v}_{\perp}$ and the spectral energy
density of Langmuir waves $W(v,x,t)=\int W(\vec{k}) d\vec{k}_{\perp}$ evolve
following the non-linearly coupled kinetic equations:
\begin{equation} \label{eq:kin1}
  \frac{\partial f}{\partial t}+v\frac{\partial f}{\partial x}=
  \frac{4\pi^2 e^2}{m^2}\frac{\partial
  }{\partial v}\frac{W}{v}\frac{\partial f}{\partial v}=\frac{\partial
  }{\partial v}D \frac{\partial f}{\partial v}\,,
\end{equation}
\begin{equation}\label{eq:kin2}
  \frac{\partial W}{\partial t}=\frac{\pi \omega _{pe}}{n_p}v^{2}W
  \frac{\partial f}{\partial v}\,,
\end{equation}
where $\int Wdk=U$ and $\int f\text{d}v=n_b$ are the energy density of Langmuir
waves and the number density of the electron beam. For completeness, we note
that the spontaneous terms are not taken into account in the kinetic model
(equations \ref{eq:kin1},\ref{eq:kin2}),
since the beam-driven level of Langmuir waves is much
higher than the spontaneous/thermal one
\citep[see discussion in][]{2017SoPh..292..117L}.
Equation (\ref{eq:kin2}) does not include the spatial transfer of the energy by
Langmuir waves, since the group velocity of Langmuir waves is small ($v_{gr}\sim
v^2_{T_e}/v\ll v$, where $v_{T_e}$ is the electron thermal velocity). The
kinetic equations (\ref{eq:kin1}-\ref{eq:kin2}) do not have an analytical
solution and additional assumptions are required to solve this system of
equations. For completeness, we note that the equations (\ref{eq:kin1},\ref{eq:kin2})
are coupled to nonlinear processes responsible for decay/coalescence
of Langmuir waves. The wave-wave interactions are normally treated numerically,
see e.g. the large-scale simulations by \citet{2014A&A...572A.111R}.

\section{Hydrodynamic description}\label{sec:hydrodinamic description}

The characteristic time of beam-plasma interaction is normally small
$\tau_\text{q}\ll t=d/v$, where $d$ is the size of an electron beam.
The smallness of quasilinear time
allows using hydrodynamic description for beam electrons and Langmuir waves
\citep{1970JETP...31..396R,1995PlPhR..21...89M,1999SoPh..184..353M,2018PhPl...25j0501R},
so the
electron distribution function $f(v,x,t)$ in the kinetic equations
(\ref{eq:kin1},\ref{eq:kin2}) is the series in small parameter
$\tau_\text{q}v/d$
\begin{equation}\label{eq:powers}
  f=f^{0}+f^{1}+...\,,
\end{equation}
which is conceptually similar to the Chapman-Enskog theory of a neutral gas
dominated by collisions \citep{1970mtnu.book.....C}. Unlike the Chapman-Enskog
theory, this theory utilises fast beam-plasma interaction via Langmuir waves.
Substituting the expansion (\ref{eq:powers}) into kinetic equations
(\ref{eq:kin1},\ref{eq:kin2}), we have in $0^{\text{th}}$-order or the fastest
terms when $\tau_\text{q}v/d \rightarrow 0$:
\begin{align}
  0 =& \frac{4\pi^2 e^2}{m^2}\frac{\partial
}{\partial v}\frac{W^0}{v}\frac{\partial f^0}{\partial v} \propto \tau_{q}^{-1}, \label{eq:zero}\\
  0= & \frac{\pi \omega _{pe}}{n_{p}}v^{2}W^0
\frac{\partial f^0}{\partial v} \propto \tau_\text{q}^{-1}. \label{eq:zero2}
\end{align}
and hence dominant for $d/v \gg \tau_\text{q}$. This leads to a well-known
result that the $0^{\text{th}}$-order solution is a plateau in the velocity
space since ${\partial f^0}/{\partial v}=0$ \citep[e.g.][]{1967PlPh....9..719V}
\begin{equation}\label{eq:plateau}
  f^{0}\left(v,x,t\right) =\left\{
\begin{array}{ll}
p\left( x,t\right) , & 0<v<u(x,t) \\
0, &v\geq u(x,t)%
\end{array}%
\right.
\end{equation}
and an enhanced level of Langmuir waves, so that the spectral energy density of
Langmuir waves becomes
\begin{equation}
W^{0}\left(v,x,t\right) =\left\{
\begin{array}{ll}
W_{0}\left(v,x,t\right) , &0<v<u(x,t) \\
0,&v\geq u(x,t)%
\end{array}%
\right.
\end{equation}%
where zero-order terms $f^{0}$ and $W^{0}$ turn the right-hand sides of kinetic equations
(\ref{eq:kin1}, \ref{eq:kin2}) to zero. In other words, any initially unstable
electron distribution function relaxes to a plateau and Langmuir waves are
generated within quasilinear time $\sim \tau_\text{q}$. Here $p\left(
x,t\right)$ is the plateau height and $u(x,t)$ is the maximum electron velocity,
so that the number density of electrons is
\begin{equation}\label{eq:p_norm}
  n(x,t)=\int\limits_{0}^{u(x,t)}p\left( x,t\right) \text{d}v=p\left( x,t\right) u(x,t).
\end{equation}
Following \citep{1999SoPh..184..353M}, one can find the equations for $p(x,t)$,
$u(x,t)$ and $W_0(v,x,t)$. Integrating equation (\ref{eq:kin1}) over $v$ from
$v=0$ to $v=u(x,t)$, one obtains the equation for electron number density
$n(x,t)=p(x,t)u(x,t)$
\begin{equation}\label{eq:1stHD}
  \frac{\partial pu}{\partial t}+\frac{1}{2}\frac{\partial pu^{2}}{\partial x}
  =\frac{\partial n}{\partial t}+\frac{1}{2}\frac{\partial nu}{\partial x}=0,
\end{equation}
which is the hydrodynamic continuity equation or conservation of electrons.
Integrating equation (\ref{eq:kin1}) over $v$ between $u-\xi$ and $v=u+\xi$,
with $\xi \xrightarrow[]{} 0$, gives
 \begin{equation}\label{eq:2ndHD}
  \frac{\partial u}{\partial t}+u\frac{\partial u}{\partial x}=0,
\end{equation}
while combining equations (\ref{eq:kin1},\ref{eq:kin2}) one obtains
\begin{equation}\label{eq:eqW0}
  \frac{\partial p}{\partial t}+v\frac{\partial p}{\partial x}=
\frac{\omega_{pe}}{m}\frac{\partial
}{\partial v}\frac{1}{v^3}\frac{\partial W_0}{\partial t}\,,
\end{equation}
which is the equation for the spectral energy density of Langmuir waves.
Equation (\ref{eq:eqW0}) can be integrated to find a solution for a initial
value problem. Following \citet{1995PlPhR..21...89M,1999SoPh..184..353M}, the
equations for $p(x,t)$, $u(x,t)$ and $W(v,x,t)$ can be integrated for given
initial conditions. For initial condition
\begin{equation}\label{eq:ft0}
  f(v,x,t=0)=n_b \exp(-x^2/d^2) g(v),
\end{equation}
where $n_b$ is the electron beam density at $x=0$, and $g(v)=2v/v_0^2$ for
$v<v_0$, the solution of equations (\ref{eq:1stHD},
\ref{eq:2ndHD},\ref{eq:eqW0}) gives (see \citep[e.g.][]{2001CoPhC.138..222K})
\begin{align}
  &u(x,t) =  v_0 \,, \label{eq:solutionHD1}\\
  &p(x,t) = \frac{n_b}{v_0} \exp(-(x-v_0t/2)^2/d^2) \,,\label{eq:solutionHD2}\\
  &W_0(v,x,t)= \frac{m}{\omega_{pe}}v^4\left(1-\frac{v}{v_0}\right)p(x,t)\,,\label{eq:solutionHD3}
\end{align}
The solution (\ref{eq:solutionHD1}-\ref{eq:solutionHD3}) suggests a
beam-Langmuir-wave structure, i.e. electron beam together with Langmuir waves
propagate with speed $v_0/2$ preserving the initial size $d$. The equations as
well as solution assume that the relaxation proceeds at the same rate for all
$x$ and $t$, which is evidently not true due to finite spatial size of the
electron beam $d$. The electron number density is higher near the peak of the
beam-plasma structure and decreases away. Therefore, the relaxation of electrons
should proceed at different rate $\tau _\text{q}^{-1}\propto n(x,t)$ in various
spatial locations and one should take into account the spatial variation of
$\tau_\text{q}$ due to variation of electron number density $n(x,t)$ of the
beam.

\section{Hydrodynamics with non-linear diffusion}\label{sec:nonlin_diff}

To address the inhomogeneity of quasilinear time, we retain the $f^{1}$ term in
the expansion of $f$ and substituting (\ref{eq:powers}) into (\ref{eq:kin1}),
one finds
\begin{equation}\label{eq:exp_eq}
\frac{\partial \left( f^{0}+f^{1}\right) }{\partial t}+v\frac{\partial
\left( f^{0}+f^{1}\right) }{\partial x}=
 \frac{\partial p}{\partial t}+v\frac{\partial p}{\partial x}+\frac{\partial
f^{1}}{\partial t}+v\frac{\partial f^{1}}{\partial x}=
\frac{\partial }{\partial v}D\frac{%
\partial f^{1}}{\partial v}  \,,
\end{equation}
where the first order terms are retained. Further, integrating this equation
over velocity from $0$ to $\infty$ gives
\begin{equation}\label{eq:f_int}
\frac{\partial }{\partial t}\int\limits_{0}^{v_0}\left( f^{0}+f^{1}\right)
\text{d}v+\frac{\partial }{\partial x}\int\limits_{0}^{v_0}
v\left(f^{0}+f^{1}\right) \text{d}v=\left. D\frac{\partial f^{1}}{\partial v}\right\vert
_{0}^{v_0}\,,
\end{equation}
where due to the quasilinear relaxation at $t\gg \tau_\text{q}$, a plateau is
considered to be established in the electron distribution function i.e.
$f^0(v,x,t) = p(x,t)$ for $v<v_0$. Hence, equation~(\ref{eq:f_int}) becomes
\begin{equation}\label{eq:kinetic_diff}
\frac{\partial pv_0}{\partial t}+\frac{1}{2}\frac{\partial pv_0^{2}}{%
\partial x}+\frac{\partial }{\partial t}\int\limits_{0}^{v_0}f^{1}\text{d}v+\frac{%
\partial }{\partial x}\int\limits_{0}^{v_0}vf^{1}\text{d}v=0\,,
\end{equation}
where $\int\limits_{0}^{v_0}f^{1}\text{d}v=0$ because
$\int\limits_{0}^{v_0}(f^{0}+f^{1})\text{d}v=n(x,t)$. Since the electron number
density can be written $n(x,t)=p(x,t)v_0$, we can write equation
(\ref{eq:kinetic_diff}) as
\begin{equation}\label{eq:f1_eq}
\frac{\partial n}{\partial t}+\frac{v_0}{2}\frac{\partial n}{\partial x}+%
\frac{\partial }{\partial x}\int\limits_{0}^{v_0}vf^{1}\text{d}v=0\,,
\end{equation}%
where $f^{1}(v,x,t)$ is to be found. The procedure to find $f^1$ is similar to the derivation
of a spatial diffusion coefficient from pitch-angle scattering diffusion
coefficient \citep{1966ApJ...146..480J,1970ApJ...162.1049H,1989ApJ...336..243S}.
Multiplying equation (\ref{eq:exp_eq}) by $v_0$ and subtracting equation
(\ref{eq:f1_eq}) one finds
\begin{equation}\label{eq:diff_deriv}
  \left( v-\frac{v_0}{2}\right) \frac{\partial n}{\partial x}+v_0\frac{%
    \partial f^{1}}{\partial t}+v_0v\frac{\partial f^{1}}{\partial x}-\frac{%
    \partial }{\partial x}\int\limits_{0}^{v_0}vf^{1}dV=v_0\frac{\partial }{%
    \partial v}D\frac{\partial f^{1}}{\partial v}\,,
\end{equation}%
and retaining only zero order terms, one finds equation for $f^1(v,x,t)$
\begin{equation}\label{eq:diff_deriv2}
\frac{1}{v_0}\left( v-\frac{v_0}{2}\right) \frac{\partial n}{\partial x}=%
\frac{\partial }{\partial v}D\frac{\partial f^{1}}{\partial v}\,,
\end{equation}%
where the right-hand side is zero order due to fast plateau formation (equation
\ref{eq:zero}). Since the quasilinear relaxation operates $0<v<v_0$, velocity
diffusion coefficient $D=\pi\omega_{pe}^2/(mn_p)(W/v)$ [One can see this
explicitly from $W_0$ solution (\ref{eq:solutionHD3})] should be zero at the
boundary velocities, i.e.:
\begin{equation}
    \left. D\right\vert _{v=0}=\left. D\right\vert _{v=v_0}=0.
\end{equation}
These boundary conditions allow us to integrate equation (\ref{eq:diff_deriv2})
over $v$ and obtain
\begin{equation}
D\frac{\partial f^{1}}{\partial v}=\frac{1}{v_0}\frac{\partial n}{\partial
x}\int\limits_{0}^{v}\left( v^{^{\prime }}-\frac{v_0}{2}\right)
\text{d}v^{^{\prime }}=\frac{1}{v_0}\frac{\partial n}{\partial x}\frac{1}{2}%
v\left( v-v_0\right) +C_{1}\,,
\end{equation}%
where $C_{1}=0$ due to $\left. D\right\vert _{0}=\left. D\right\vert _{v_0}=0$,
yielding
\begin{equation}\label{eq:df1_dv}
\frac{\partial f^{1}}{\partial v}=\frac{v\left( v-v_0\right) }{2v_0D}%
\frac{\partial n}{\partial x}.
\end{equation}%
Further, integrating equation (\ref{eq:df1_dv}) over $v$, we find expression for $f^1$
\begin{equation}\label{eq:f_1_c2}
f^{1}=\frac{1}{2v_0}\frac{\partial n}{\partial x}\int\limits_{0}^{v}\frac{%
v^{^{\prime }2}-v_0v^{^{\prime }}}{D}\text{d}v^{^{\prime }}+C_{2}\,,
\end{equation}
where the constant $C_2$ is determined from
$\int\limits_{0}^{v_0}f^{1}\text{d}v=0$, (see Appendix \ref{appendix:diffusion}
for details). Therefore, $f^1$ becomes
\begin{equation}
f^{1}=\frac{1}{2v_0}\frac{\partial n}{\partial x}\left[ \int\limits_{0}^{v}%
\frac{v^{^{\prime }2}-v_0v^{^{\prime }}}{D}\text{d}v^{^{\prime }}-\frac{1}{v_0}%
\int\limits_{0}^{v_0}
\left( v_0-v^\prime \right) \frac{v^{\prime 2}-v_0v^\prime}{D}\text{d}v^\prime\right] \,,
\end{equation}
and from equation (\ref{eq:vf1_int})
\begin{equation}
\int\limits_{0}^{v_0}vf^{1}\text{d}v=- \frac{1}{4v_0}\frac{\partial n}{\partial x}\int\limits_{0}^{v_0}%
\frac{v^{2}\left( v_0-v\right) ^{2}}{D}\text{d}v\,.
\end{equation}
Hence, the transport equation for electron number density
(equation~\ref{eq:f1_eq}) takes the form
\begin{equation}\label{eq:n_diff_f1}
\frac{\partial n}{\partial t}+\frac{v_0}{2}\frac{\partial n}{\partial x}
=-\frac{\partial }{\partial x}\int\limits_{0}^{v_0}vf^{1}\text{d}v
=\frac{\partial }{\partial x}\frac{1}{4v_0}\frac{\partial n}{\partial x}\int\limits_{0}^{v_0}%
\frac{v^{2}\left( v_0-v\right) ^{2}}{D}\text{d}v\,,
\end{equation}
which is the modified equation of particle conservation
(compare to equation~\ref{eq:1stHD}).

\subsection{Advection and non-linear diffusion}

The velocity diffusion coefficient $D\propto W$ is determined by the level of
Langmuir waves. Taking the spectral energy density of Langmuir waves $W^0$ given
by equation (\ref{eq:solutionHD3}), one can write for $D$:
\begin{equation}\label{eq:D_coef}
D=\pi \frac{\omega _{pe}}{v_0}\frac{n(x,t)}{n_{p}}
v^{3}\left( 1-\frac{v}{v_0}\right)
=D_{0}v^{3}\left( 1-\frac{v}{v_0}\right)\,,
\end{equation}%
where $D_0=\pi \frac{\omega _{pe}}{v_0}\frac{n(x,t)}{n_{p}}$. Finally
substituting (\ref{eq:D_coef}) into (\ref{eq:n_diff_f1}) and integrating from
$v_{\min}$ (instead of $0$ as in equation \ref{eq:n_diff_f1}) leads us to
advection diffusion equation
\begin{equation} \label{eq:v_0_diffus_eq}
\frac{\partial n}{\partial t}+\frac{v_0}{2}\frac{\partial n}{\partial x}-
\frac{\partial}{\partial x} D_{xx} \frac{\partial n}{\partial x}=0,
\end{equation}
with the non-linear spatial diffusion coefficient given by
\begin{equation}\label{eq:D_xx}
    D_{xx}=\frac{1}{4v_0}\int\limits_{v_{\min}}^{v_0}%
\frac{v^{2}\left( v_0-v\right) ^{2}}{D}\text{d}v= \frac{v_0^2 n_{p}}{4 \pi \omega _{pe} n(x,t)}\left( \ln \frac{v_0}{v_{\min }}-1\right) \propto \frac{v_0^2}{4 }\tau_\text{q}\,.
\end{equation}
The spatial diffusion coefficient (\ref{eq:D_xx}) is dependent on the beam
density and is smaller for smaller quasilinear time. In other words, the
stronger the beam (larger $n_b$), the slower the diffusion term (smaller
$D_{xx}$). The diffusion coefficient with $1/n(x,t)$ nonlinearity is often
named \textit{fast diffusion} i.e. diffusion is fast in regions where the
density of particles is low
\citep[e.g.][]{doi:10.1080/03605308808820566,2017arXiv170608241V}.

An important peculiarity of the solution (\ref{eq:D_xx}) is that we had to
integrate from $v_{\min}$ and when $v_{\min}\rightarrow 0$, $D_{xx}\rightarrow
\infty$ diverges. The divergence has a physical reason: the plateau down to zero
velocity is formed at $t \rightarrow \infty$, so the diffusion
coefficient is infinite when $v_{\min}\rightarrow 0$. In other words,
the spatial diffusion coefficient, $D_{xx}\propto \tau_\text{q}$, is infinite
due to the infinite time required to form plateau down to $v_{\text{min}}=0$.
While the plateau is quickly formed
over broad range of velocities (over quasilinear time $\tau_\text{q}$), the
growth rate of Langmuir waves is actually zero at $v=0$, as one can see from
equation (\ref{eq:kin2}).  Indeed, numerical simulations
\citep[e.g.][]{1998PlPhR..24..772K,2001CoPhC.138..222K} show that the relaxation
proceeds down to a small but finite velocity. In plasma with a Maxwellian
distribution of thermal particles, the plateau is also formed between maximum
beam velocity and thermal distribution, so that $v_{\text{min}}\simeq
3-4v_{T_e}$
\citep[e.g.][]{2002PhRvE..65f6408K,2008PhPl...15c2303Z,2011PPCF...53h5004Z,2019JGRA..124...68S}.

Therefore, in order to compare our analytical model to the results of the
numerical simulations or observations, we include a constant lower bound
$v_{\text{min}}$ to the plateau in velocity, resulting in the new electron
distribution function
\begin{equation}\label{eq:plateau_vmin}
  f^{0}\left(v,x,t\right) =\left\{
\begin{array}{ll}
p\left( x,t\right) , & v_{\text{min}}<v<u(x,t) \\
0, & v\leq v_{\text{min}}, \ v\geq u(x,t)%
\end{array}%
\right.
\end{equation}
and the spectral energy density of Langmuir waves
\begin{equation}
W^{0}\left(v,x,t\right) =\left\{
\begin{array}{ll}
W_{0}\left(v,x,t\right) , &v_\text{min}<v<u(x,t) \\
0,& v\leq v_{\text{min}}, \ v\geq u(x,t)
\end{array}%
\right.
\end{equation}%
where the solution for $u(x,t)$, $p(x,t)$, and $W_0(x,v,t)$ can be found
following \citet{1998PlPhR..24..772K} to be
\begin{align}
&u(x,t)=v_0, \label{eq:sol_vmin1}\\
&p(x, t)=\frac{n_b}{v_0-v_{\text {min}}} \exp \left[-\frac{\left(x-\left(v_0+v_{\text {min}}\right) t / 2\right)^2}{d^2}\right], \label{eq:sol_vmin2}\\
&W_{0}\left(v,x,t\right) =\frac{m}{\omega_{pe}} v^3\left(v-v_{\min }\right)\left(1-\frac{v+v_{\min }}{v_0+v_{\min }}\right) p(x, t) \label{eq:sol_vmin3},
\end{align}
The obvious difference from the solution
(\ref{eq:solutionHD1}-\ref{eq:solutionHD3}) is that the solution
(\ref{eq:sol_vmin1}-\ref{eq:sol_vmin3}) accounts for the minimum velocity of a
plateau. Another consequence of $v_{\min}$ is that electron density is now given
by
\begin{equation}\label{eq:p_norm_vmin}
  n(x,t)=\int\limits_{v_{\text{min}}}^{v_0}p\left( x,t\right) \text{d}v=p\left( x,t\right) (v_0-v_{\text{min}}) \,\, ,
\end{equation}
with a new diffusion equation
\begin{equation}\label{eq:n_f1_eq_vmin}
\frac{\partial n}{\partial t}+\frac{v_0 + v_{\text{min}}}{2}\frac{\partial n}{\partial x}+
\frac{\partial }{\partial x}\int\limits_{v_{\min}}^{v_0}vf^{1}\text{d}v=0 \,\,.
\end{equation}
After finding $f^1$ (see Appendix~(\ref{app:non-0_low_bound})), we arrive to

\begin{equation}
\int\limits_{v_{\text{min}}}^{v_0}vf^{1}\text{d}v=-\frac{1}{4(v_0-v_{\text{min}})}\frac{\partial n}{\partial x}\int\limits_{v_{\text{min}}}^{v_0}%
\frac{(v-v_{\text{min}})^{2}\left( v_0-v\right) ^{2}}{D}\text{d}v \,\,.
\label{intfastdiff}
\end{equation}

Similarly to the previous subsection, the expression for $D$ can be found from
the formula for the spectral energy density $W_0(x,v,t)$ and the plateau height
$p(x,t)$. From \citet{1998PlPhR..24..772K}, we have that in case of non-zero
$v_{\min}$
\begin{equation}\label{eq:D_v_vmin}
    D=D_{0}v^2\left(v-v_{\min }\right)\left(1-\frac{v+v_{\min }}{v_0+v_{\min }}\right) ,
\end{equation}
where now $D_{0}=\pi \frac{\omega _{pe}}{(v_0-v_{\text{min
}})}\frac{n(x,t)}{n_{p}}.$

If we insert this $D$ into equation (\ref{intfastdiff}), we obtain (see
Appendix~\ref{app:non-0_low_bound})
\begin{equation}\label{eq:int_vf1_vmin}
    \int\limits_{v_{\min }}^{v_0}vf_{1}\text{d}v=-\frac{v_0+v_{\text{min}}}{4D_{0}}\left( \frac{v_0+v_{\text{min}}}{v_0-v_{\text{min}}}\ln
\frac{v_0}{v_{\min }}-2\right)\frac{\partial n}{\partial x},
\end{equation}
and our advection-nonlinear-diffusion becomes
\begin{equation} \label{eq:advection_diff_vmin}
\frac{\partial n}{\partial t}+\frac{(v_0 + v_{\text{min}})}{2}\frac{\partial n}{\partial x}-\frac{\partial }{\partial x}D_{xx}\frac{\partial n}{\partial x}=0,
\end{equation}%
where our new diffusion coefficient $D_{xx}$ that includes $v_{\min}$ is now given by
\begin{equation}\label{eq:Dxx_vimin}
  D_{xx} = \frac{v_0^2-v_\text{min}^2}{4\pi \omega_{pe}}\frac{n_p}{n(x,t)}\left(
\left(\frac{v_0+v_{\text{min}}}{v_0-v_{\text{min}}}\right)\ln
\frac{v_0}{v_{\min }}-2\right)= \frac{v_0^2-v_{\text{min}}^2}{4\pi}\tau_q\left(
\left(\frac{v_0+v_{\text{min}}}{v_0-v_{\text{min}}}\right)\ln
\frac{v_0}{v_{\min }}-2\right)\,.
\end{equation}
The spatial diffusion coefficient $D_{xx}$ is inversely proportional to electron
number density $n(x,t)$ or is proportional to quasilinear time, so the spatial
diffusion is faster for longer quasilinear time $n_p/(\omega_{pe}n(x,t))$. The
diffusion coefficient $D_{xx}$ is zero when $v_{\min }=v_0$, i.e.
spatial diffusion is not possible without quasilinear relaxation.
The electron beam diffusion coefficient (equation~\ref{eq:Dxx_vimin})
is also dependent on $v_{\min}$, so the spatial expansion of electron beam
is larger for smaller $v_{\min}$.

\section{Asymptotic solution to advection-nonlinear-diffusion equation}

Let us consider the evolution of an electron beam given by initial condition
\begin{equation}\label{eq:init_delta}
   n(x,t=0)=n_b\delta \left( x/d\right)\,,
\end{equation}
where $n_b$ is the electron beam density and $d$ is the characteristic size. The
advection-nonlinear diffusion equation (\ref{eq:advection_diff_vmin}) with
$n(x,t)$ normalised with $n_b$ can be rewritten
\begin{equation}\label{eq:gendiffeq}
    \frac{\partial n}{\partial t}+\frac{v_0+v_{min}}{2}\frac{\partial n}{\partial x}-
    \frac{\partial}{\partial x}D_{xx}^0\frac{n_{b}}{n}\frac{\partial n}{\partial x}=0\,,
\end{equation}
where the nonlinear dependency of $D_{xx}$ on $n(x,t)$ is explicitly highlighted
by introducing $D_{xx}=D_{xx}^0\frac{n_{b}}{n}$. The equation
(\ref{eq:gendiffeq}) can be solved for constant $v_0$ and $v_{\min}$ to find
asymptotic solution
\begin{equation}\label{eq:lorentz_sol}
\begin{aligned}
     n(x,t)=\left(\frac{\left(x-{(v_0+v_{\min})}t/{2}\right)^2}{2D^0_{xx}n_{b}t} +
    \frac{2\pi^2}{n_bd^2}D^0_{xx}t\right)^{-1}=\\
    =\frac{n_b}{\pi} \frac{2\pi D^0_{xx}t/d^2}{(x-(v_0+v_{\min})t/{2})^2/d^2+ 4\pi^2 (D^0_{xx}t)^2/d^4},
\end{aligned}
\end{equation}
which is a Lorentzian
\begin{equation}
 L(x')=\frac{1}{\pi}\frac{\gamma}{x'^2+\gamma^2}\,,
\end{equation}
where $x'=(x-(v_0+v_{\min})t/{2})/d$ and $\gamma=2\pi D^0_{xx}t/d^2$. The
solution (\ref{eq:lorentz_sol}) describes the expanding electron beam moving with the speed
$(v_0+v_{\min})/2$. The electron beam size given by $\gamma$ is proportional to
time $t$, and for $t\rightarrow 0$, $n(x,t)\rightarrow n_bd\delta(x)$, which is
the initial condition (\ref{eq:init_delta}). The electron density equation
(\ref{eq:gendiffeq}) without advection term has been studied for different
applications
\citep[e.g.][]{1976PhLA...59..285L,Berryman82,Esteban88,Hill93,1993RSPTA.343..337K,1995PhRvL..74.1056R,2005PhRvE..72c1106P}
and the asymptotic profiles are often referred as ZKB profiles [from Zeldovich,
Kompanyeets and Barenblat \citep[see][ as a review]{1996sssi.book.....B}].
$n(x,t)$ given by (\ref{eq:lorentz_sol}) also conserves the number of particles
$\int^{+\infty}_{-\infty}n(x,t)\text{d}x=d n_b $ as expected.

The solution (\ref{eq:lorentz_sol}) shows that the electron beam always expands
with time or with distance, so the peak of the beam at $x=(v_0+v_{\min})t/{2}$
decreases following:
\begin{equation}\label{eq:n_x_peak}
    n\left(x,t=\frac{2x}{v_0 + v_{\min}}\right)
    = \frac{n_b d}{\pi}\frac{(v_0 + v_{\min})d}{4\pi xD^0_{xx}}\propto \frac{d}{x}\,,
\end{equation}
which is an important result for the theory of type III bursts. The plausible
decrease of electron number density with distance $1/x$ would be preferable to
explain the intensity of type III burst with distance observed in the
interplanetary space \citep[e.g.][]{2014SoPh..289.3121K}.

Another interesting consequence of the solution (\ref{eq:lorentz_sol}) is that
the beam size is growing with time or distance. Full Width at Half Maximum
(FWHM) of the electron beam $ \Delta x$ is
\begin{equation}\label{eq:FWHM_x}
    \Delta x=2\gamma d
    =\frac{4\pi}{d} D^0_{xx}t
    =\frac{4\pi}{d} D^0_{xx}\frac{2x}{(v_0+v_{\min})}
    \propto  \tau_\text{q} \frac{x}{d}\,,
\end{equation}
or dividing by the average speed of the electron beam $(v_0+v_{\min})/{2}$, one
obtains the time FWHM of the beam
\begin{equation}\label{eq:FWHM_t}
    \Delta t=\frac{4\gamma d}{v_0+v_{\min}}
    =\frac{4\pi}{d} D^0_{xx} \frac{2t}{(v_0+v_{\min})}
    =\frac{4\pi}{d} D^0_{xx} \frac{4x}{(v_0+v_{\min})^2}
    \propto \tau_\text{q}\frac{x}{d}\,,
\end{equation}
where the spatial expansion of the beam $\Delta x$ is linearly growing with
time $\Delta x \propto t$ or the particle propagate ballistically, which is the
special case of super-diffusion. Constant spatial diffusion coeficient leads to
$\Delta x \propto t^{1/2}$, but the non-linear diffusion $D_{xx} \propto
1/n(x,t)$ due to Langmuir wave turbulence make electron beam to expand
"faster", i.e.  $\Delta x \propto t$ which is so-called super-diffusion
\citep[e.g.][]{1984PhLA..105..169O,1997GeoRL..24.1727T,2006ApJ...639L..91Z}.
Here, the Langmuir turbulence is self-consistently generated as the electron
beam propagates and expands in space. The level of Langmuir turbulence is
proportional to the number of particles that gives the nonlinearity of the
diffusion coefficient.

\subsection{Initially finite beam dynamics}

To compare with observations and numerical simulations, let us consider initial
electron number density function as a Gaussian with characteristic size $d$,
which is similar to the initial condition in \citep{1998PlPhR..24..772K}, i.e.
the electron distribution function at $t=0$ is
\begin{equation}\label{eq:fxvt_t0}
  f(v, x, t=0)=\frac{2 n_b}{v_0}\frac{v}{v_0} \exp \left(-\frac{x^2}{d^2}\right)\,,
  \;\; 0<v<v_0\,,
\end{equation}
hence the number density of beam electrons
\begin{equation}\label{eq:gauss_init}
    n(x,t=0)=n_b \exp\left(-x^2/d^2\right)\,,
\end{equation}
with the total number of particles $\int_{-\infty}^{+\infty}n(x,t)=n_b
d\sqrt{\pi}$. Then the solution of advection-nonlinear-diffusion equation
(\ref{eq:gendiffeq}) is the convolution of the initial condition
(\ref{eq:gauss_init}) and Lorentzian from equation~(\ref{eq:lorentz_sol})
normalised to $1$, which is the solution to the Dirac delta function initial
condition (the Green's function solution, which is an approximation
when $D_{xx}$ is nonlinear,
see  e.g. \citet[][]{Kheifets1982,2008IJMPA..23..299F,Frank2009}):
\begin{equation} \label{eq:gauss_sol}
    \begin{aligned}
n(x,t)  =  \frac{n_b}{\pi} \int_{-\infty} ^{\infty}
    \frac{ 2\pi D^0_{xx}t/d e^{-s^2/d^2} d s}{(x-s-\frac{v_0+v_{\min}}{2}t)^2+
    4\pi^2 (D^0_{xx}t)^2/d^2}= \\
= n_b \frac{\gamma }{\pi } \int_{-\infty} ^{\infty} \frac{e^{-y^2} dy}{(\eta - y)^2 + \gamma^2}= n_b V(\gamma(t), \eta(x,t)),
    \end{aligned}
\end{equation}
where $\eta(x,t)=(x-(v_0+v_{\min})t/2)/d$, $\;\gamma(t)={2\pi D^0_{xx}t}/{d^2}$,
$\;y=s/d$, and
\begin{equation}\label{eq:voigt}
    V(\gamma, \eta)\equiv \frac{\gamma}{\pi} \int_{-\infty}^{\infty}
    \frac{e^{-y^2} d y}{\gamma^2+(\eta-y)^2}\,,
\end{equation}
is the Voigt profile \citep{1970hmfw.book.....A}, which is the convolution of
Gaussian and Lorentzian, often used to fit spectral lines
\citep[e.g.][]{2016A&A...590A..99J}.

In case of the solution (\ref{eq:gauss_sol}), the width of the electron beam is
the combination of Lorentzian FWHM given by equation~(\ref{eq:FWHM_x}) and the
FWHM of the Gaussian (\ref{eq:gauss_init}), which is $\Delta x_G=2\sqrt{\ln2}
d\simeq 1.67 d$. The Voigt profile can be approximated \citep{WHITING19681379}:
\begin{equation}\label{eq:voigt_width}
    \Delta x_{\mathrm{V}} \approx \Delta x / 2
    +\sqrt{\Delta x ^2 / 4+\Delta x_{\mathrm{G}}^2}\sim
    \sqrt{\Delta x_{\mathrm{G}}^2+\Delta x ^2}\,,
\end{equation}
which shows that the electron beam of size $d$ is expanding ballistically with
time $\Delta x \propto  t$, when $\Delta x \gg d$. The speed of the expansion
(Equation~\ref{eq:FWHM_x}) is controlled by the quasilinear time. Smaller/larger
quasilinear time leads to slower/faster spatial electron beam expansion.

\begin{figure}
    \centering
    \includegraphics[width=0.5\textwidth]{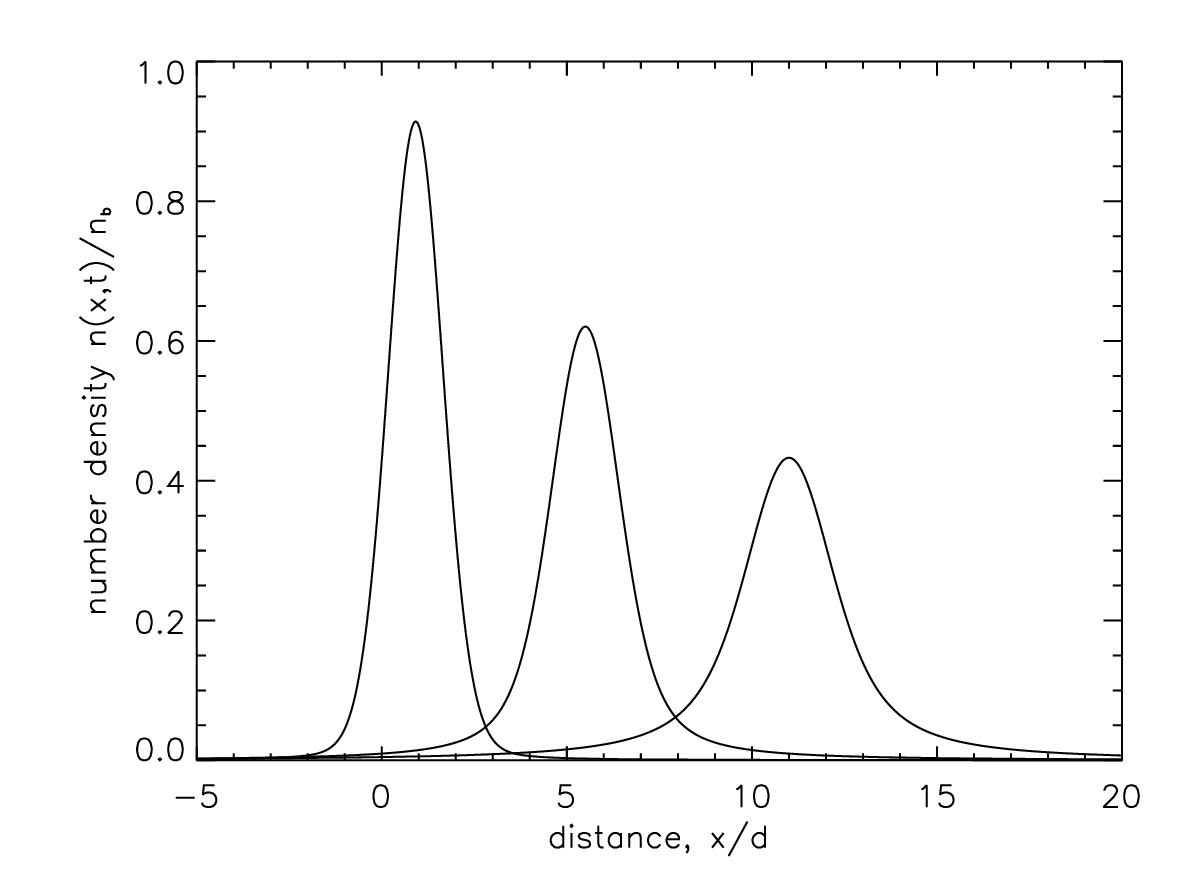}
    \caption{Electron number density profile $n(x,t)/n_b$ for the beam-plasma
    parameters as in the numerical simulations by \citet{1998PlPhR..24..772K}:
    $n_b=12$~cm$^{-3}$, $n_p=6\times 10^8$~cm$^{-3}$ (i.e. $f_{pe}\simeq
    220$~MHz) and $v_0=10^{10}$~cm/s, $v_{\min}=0.1v_0$, $d=3\times10^9$~cm. The
    three curves are the density profiles given by the solution (Equation
    \ref{eq:gauss_sol}) for $t=0.5,3,6$~seconds.}\label{fig:nxt_kontar1998}
\end{figure}

Figure~\ref{fig:nxt_kontar1998} shows the spatial evolution of electron beam
for the beam-plasma parameters used in the numerical simulations by
\citet{1998PlPhR..24..772K}. Unlike the solution assuming constant quasilinear
time (Equation~\ref{eq:solutionHD2}), the solution (Equation \ref{eq:gauss_sol})
is much closer in describing the simulated density profile showing both the
decrease of the peak density and electron beam expansion \citep[Figure 2 in
][]{1998PlPhR..24..772K}.

Using the solution for electron beam spread (\ref{eq:gauss_sol}), the spectral
energy density of the Langmuir waves (Equation \ref{eq:sol_vmin3}) becomes
\begin{equation}\label{eq:nonlin_W_solution}
W=n(x, t)\frac{m}{\omega_p}  v^3 \frac{v-v_{\min }}{v_0-v_{\min }}\left( 1-\frac{v+v_{\min }}{v_0+v_{\min }}\right)\,,
\end{equation}
so the spectral energy density decreases with distance due to the spatial evolution
of  $n(x,t)$.

\begin{figure}
    \centering
    \includegraphics[width=0.55\textwidth]{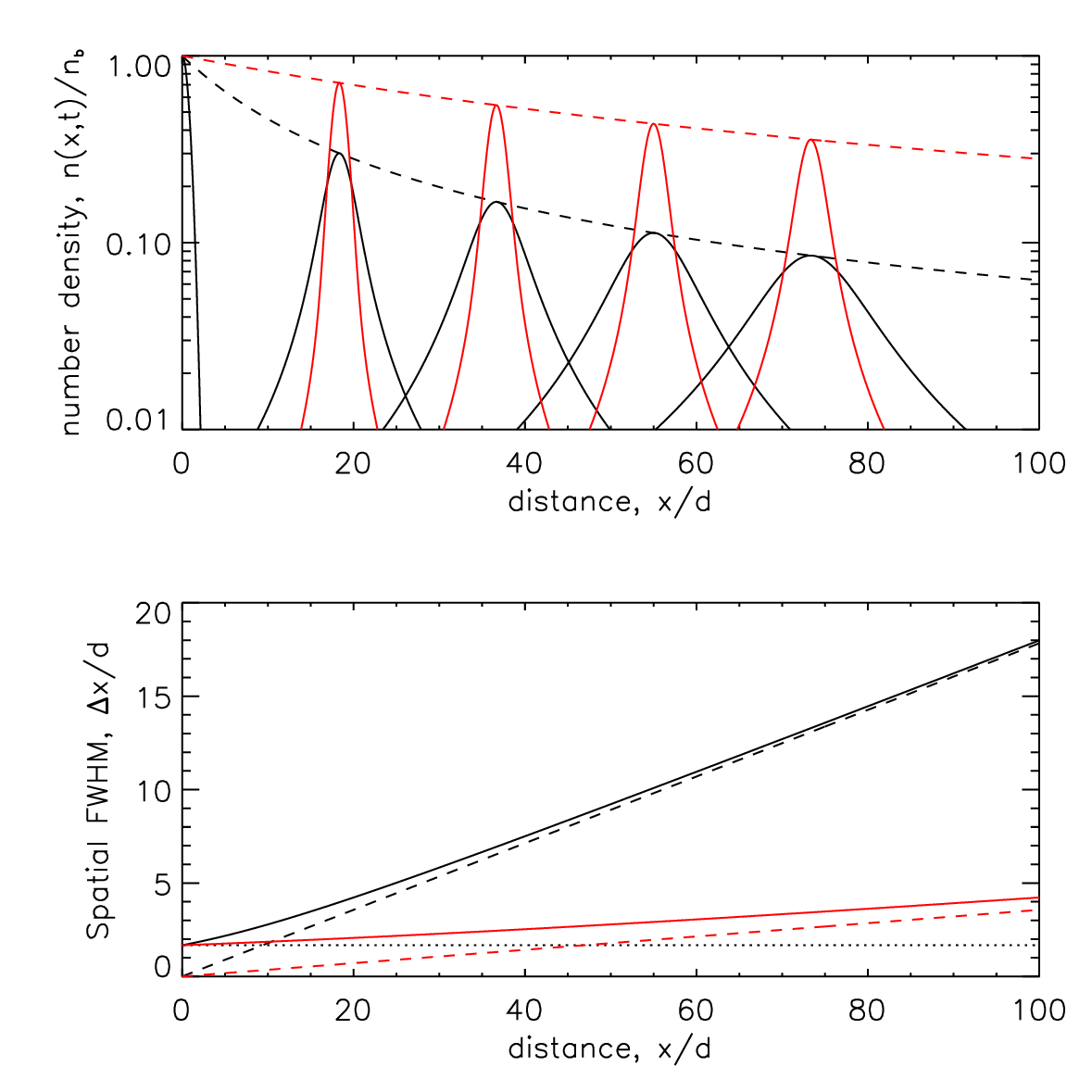}
    \caption{\textit{Top panel:}
    Electron number density profile $n(x,t)/n_b$ at $t=0,20,40,60$~seconds for
    the same beam plasma parameters as in Figure \ref{fig:nxt_kontar1998}
    $n_b=12$~cm$^{-3}$ in black and $n_b=60$~cm$^{-3}$ in red . \textit{Bottom
    panel:} FWHM width of the beam given by equation~\ref{eq:voigt_width} (solid
    line). The dashed line is the width of Lorentzian (Equation~\ref{eq:FWHM_x}).
    The horizontal dashed line is Gaussian FWHM, $\Delta x _G\simeq1.67d$. Black
    lines are for  $n_b=12$~cm$^{-3}$ and red lines are for $n_b=60$~cm$^{-3}$.}\label{fig:nxt_width}

\end{figure}

The peak density of electrons $n(x,t=2x/(v_0+v_{\min}))$ decreases with distance
following
\begin{equation}\label{eq:gauss_sol_peak}
    \frac{n\left(x,t=\frac{2x}{v_0+v_{\min}}\right)}{n_b}= V\left(\gamma\left(t=\frac{2x}{v_0+v_{\min}}\right), \eta =0\right),
\end{equation}
where $\gamma(x,t=2x/(v_0+v_{\min}))=4\pi D^0_{xx}x/((v_0+v_{\min})d^2)$. Figure
\ref{fig:nxt_width} shows the peak value (Equation \ref{eq:gauss_sol_peak}) as a
function of distance.

The width variation or the time required to pass a specific point in space for a
beam would correspond to the duration of type III burst at a given frequency.
Interestingly, type III observations, similar to the predictions of ballistic
expansion (Figure ~\ref{fig:nxt_width}), also show expansion \citep[Figure 10
in][]{2018A&A...614A..69R}. The detailed comparison would require taking into
account radio-wave propagation.  The rate of expansion is dependent on density
and can be a new valuable diagnostic of electron beam density in type III
bursts. This is also apparent from Figures~\ref{fig:new_sim_nb12} and~\ref{fig:new_sim_nb120}, where the temporal evolution of simulated electron distribution, spectral energy density and electron beam density is shown for $n_b=12$~cm$^{-3}$ and $n_b=120$~cm$^{-3}$. The value of $v_{\min}$ was chosen to be the minimum velocity value at half maximum of the electron distribution. It is evident that higher densities correspond to shorter quasilinear times of interaction, resulting in a better fit between simulations and the analytical solution.

\section{Summary}\label{discussion}

We develop a quantitative analytical model of the electron transport
responsible for type III solar radio bursts.
The developed model takes into account the
finite size of electron beam, so the generation of Langmuir waves and
quasilinear relaxation proceeds faster in the regions of higher electron number
density. In the limit of small quasilinear time, the hydrodynamic approach
yields the advection-nonlinear-diffusion equation for electron number density.
Since the rate of relaxation of electrons is governed
by the beam density at different spatial locations,
the non-linear diffusion coefficient is inversely proportional to the
beam density $D_{xx}\propto 1/n(x,t)$, process known as fast diffusion.
Low electron beam density away from the peak of the electron beam leads
to faster spatial diffusion of electrons.

\begin{figure}
\centering
   \includegraphics[width=0.9\textwidth]{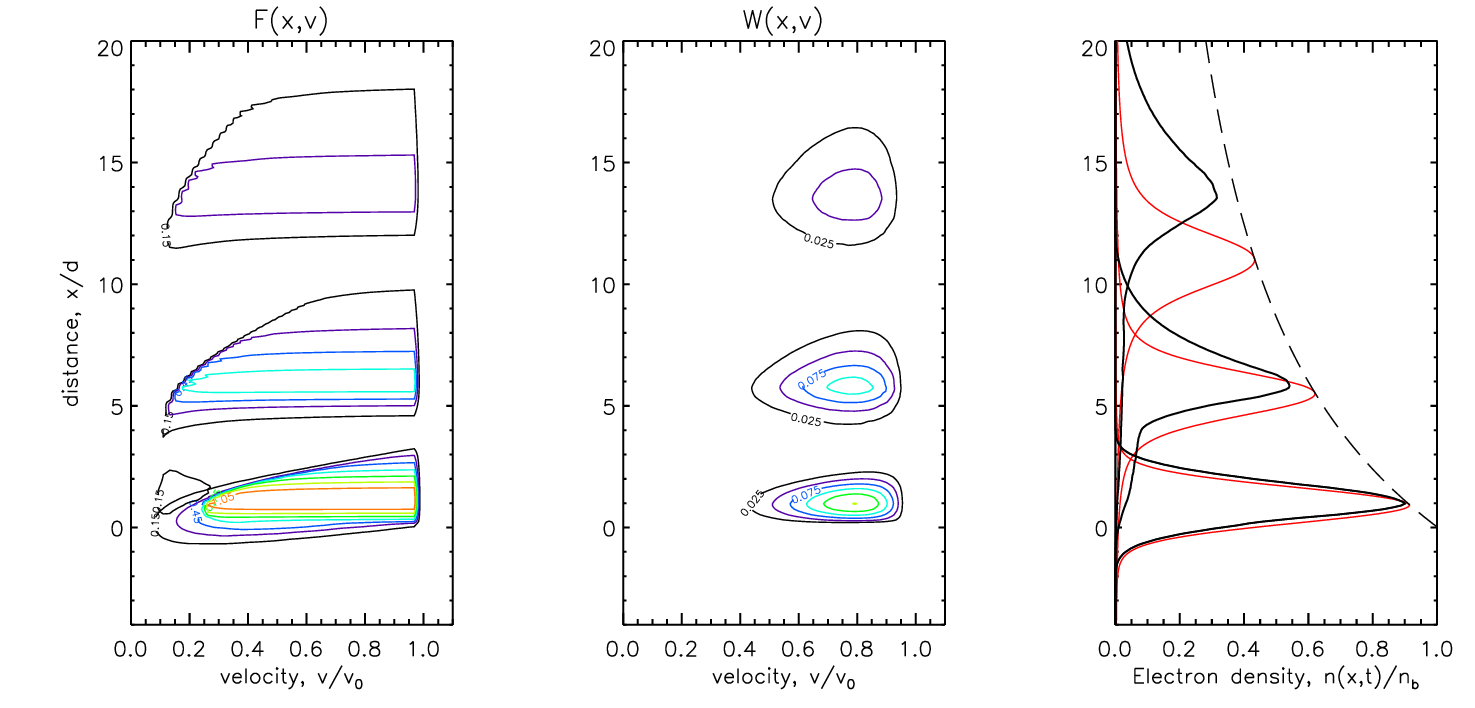}
\caption{Simulated electron distribution $f(x,v,t)$ (left), spectral energy density $W(x,v,t)$ (center), and electron beam density $n(x,t)$ (right) at three time moments $t=0.5, 3, 6$~s. for the following beam-plasma parameters $n_b=12$~cm$^{-3}$, $n_p=6\times 10^8$~cm$^{-3}$
(i.e. $f_{pe}\simeq 220$~MHz) and $v_0=10^{10}$~cm/s, $v_{\min}=0.1v_0$, $d=3\times10^9$~cm.
The analytical density profile (\ref{eq:gauss_sol}) is plotted in red, with the black dashed line showing its peak as a function of distance.} \label{fig:new_sim_nb12}
\end{figure}

\begin{figure}
\centering
   \includegraphics[width=0.9\textwidth]{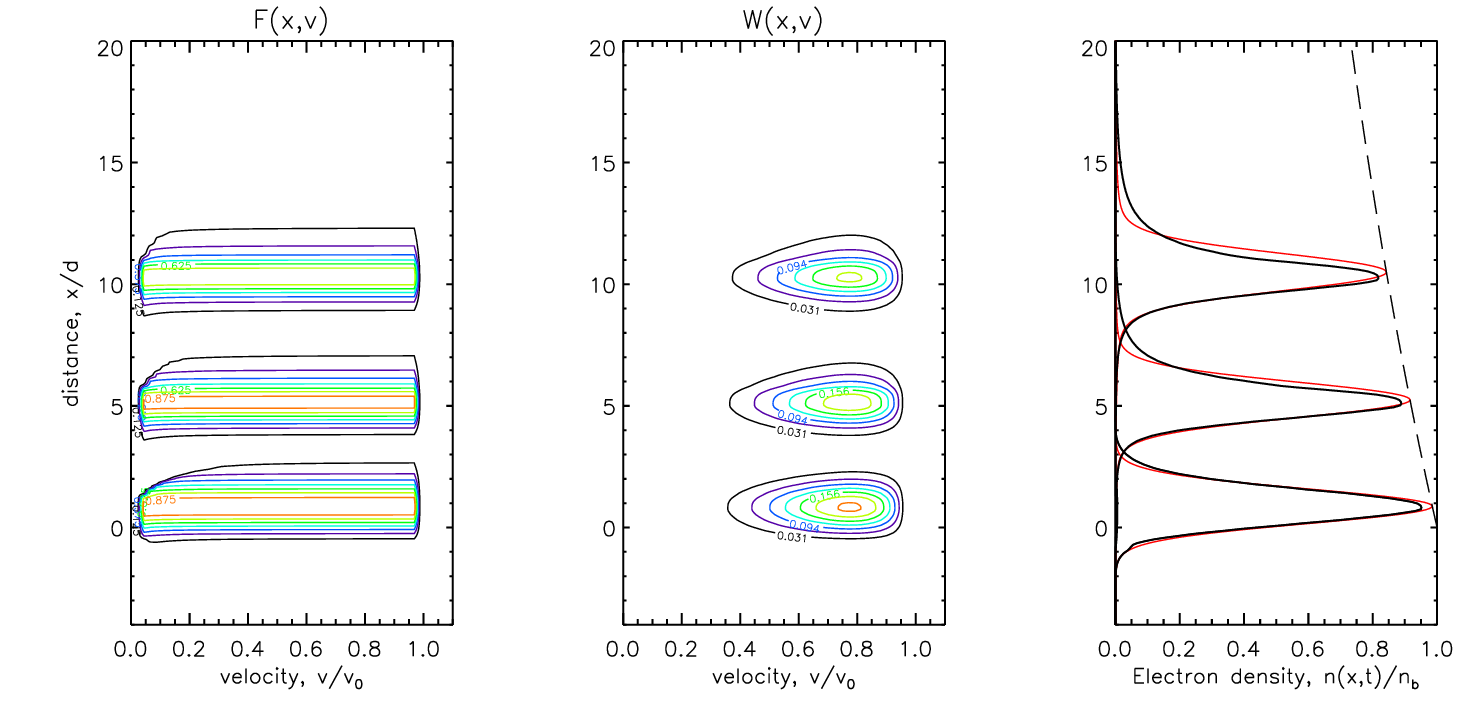}
\caption{The same as Figure \ref{fig:new_sim_nb12} but for $n_b=120$~cm$^{-3}$.}\label{fig:new_sim_nb120}
\end{figure}

The model has an elegant analytical solution showing that electron beam
propagates at constant speed $(v_0+v_{\min})/2$ but with varying spatial width.
The electron beam spatial size growths with the rate dependent
on quasilinear time $\tau_\text{q}$.
The spatial width of the electron beam is proportional to $\tau_\text{q}$
and to time $t$ at large $x\gg d$. Unlike a linear diffusion case,
when the beam size increases as $\propto \sqrt{t}$,
the nonlinear diffusion leads to ballistic (super-diffusion) expansion
i.e. electron beam size is $\propto {t}$ at large distances $x\gg d$ (see lower panel in Figure~\ref{fig:nxt_width} and equation \ref{eq:FWHM_x}).
Although the spatial expansion is linear with time,
the rate of the expansion could be small for small quasilinear times $\tau_\text{q}$
or large densities (compare Figures~\ref{fig:new_sim_nb120} and~\ref{fig:new_sim_nb12}).

The spatial expansion of the electron beam leads to the decrease
of the peak density of the electron beam. For large $x\gg d$,
when the expansion is $\propto x$, the maximum beam density decreases
as $\propto 1/x$, with the rate dependent on the beam density.
The spectral energy density of Langmuir waves as the proxy
for the type III solar radio flux is also decreasing $\propto 1/x$.

We further note that the spatial distribution of electrons has quantitative agreement with the
numerical solutions of kinetic equations, where both numerical solutions and
analytical density profiles show Voigt-like profile
(Figure~\ref{fig:nxt_kontar1998}). Similar to the simulations, the peak density
of electrons in the beam decreases with distance at the rate similar to the
numerical solution (Figure 2 in \citet{1998PlPhR..24..772K} ).
The analytical solution also shows that on top of advection with
constant speed, $(v_0+u_{\min})/2$,
the nonlinear diffusion leads to spatial expansion of electron beam with time.
The FWHM of the electron beam, in the analytical
solution is shown to be expanding ballistically, i.e. $\Delta x \propto t$ for
$\Delta x \gg d$. The expansion of the electron beam is faster
further away from the beam center due to larger local quasilinear time
since $D_{xx}\propto 1/n(x,t)$ (Equation~\ref{eq:FWHM_x}).

In application to type III solar radio bursts, the spectral energy density of
plasma emission via Langmuir waves depends on the beam density and would
decrease $\propto 1/x$, which is required to explain the radial type III solar burst flux
variations \citep{2014SoPh..289.3121K}.
The spatial expansion of the beam is also qualitatively
better fit for the time width of type III bursts \citep{2018A&A...614A..69R}.
However, more detailed studies including Langmuir wave refraction
are likely to be required to have a detailed comparison with the solar
type III burst observations.

\begin{acknowledgments}
EPK  support via the STFC/UKRI grants ST/T000422/1 and ST/Y001834/1. FA
(studentship 2604774) and EPK were supported via STFC training grant
ST/V506692/1. \end{acknowledgments}

\appendix

\section{Spatial diffusion coefficient} \label{appendix:diffusion}

To determine the constant of integration $\mathrm{C}_2$,
we note that $\int_0^{v_0} f^1 d v=0$, i.e.
\[
\int_0^{v_0} f^1 d v=0=\left.\left[v f^1\right]\right|_0 ^{v_0}-\int_0^{v_0} v \frac{\partial f^1}{\partial v} d v\,,
\]
which can be expanded into
\[
\frac{1}{2 v_0} \frac{\partial n}{\partial x} \int_0^{v_0}\left(v_0-v^{\prime}\right) \frac{v^{\prime 2}-v_0 v^{\prime}}{D} d v^{\prime}+v_0 C_2=0\,,
\]
and rearranged to have
\begin{equation}\label{eq:c2_app}
    C_2=-\frac{1}{2 v_0^2} \frac{\partial n}{\partial x} \int_0^{v_0}\left(v_0-v^{\prime}\right) \frac{v^{\prime 2}-v_0 v^{\prime}}{D} d v^{\prime}\,.
\end{equation}
Substituting $C_2$ into equation (\ref{eq:f_1_c2}), one obtains
\[
f^1=\frac{1}{2 v_0} \frac{\partial n}{\partial x}\left[\int_0^v \frac{v^{\prime 2}-v_0 v^{\prime}}{D} d v^{\prime}-\frac{1}{v_0} \int_0^{v_0}\left(v_0-v^{\prime}\right) \frac{v^{\prime 2}-v_0 v^{\prime}}{D} d v^{\prime}\right]\,,
\]
and taking into account that
\[
\int_0^{v_0} v f^1 d v=0=\left.\left[\frac{v^2}{2} f^1\right]\right|_0 ^{v_0}-\int_0^{v_0} \frac{v^2}{2} \frac{\partial f^1}{\partial v} d v\,,
\]
one finds that
\begin{equation}\label{eq:vf1_int}
    \int_0^{v_0} v f^1 d v=-\frac{1}{4 v_0} \frac{\partial n}{\partial x} \int_0^{v_0} v\left(v-v_0\right) \frac{v^2-v_0 v}{D} d v=-\frac{1}{4 v_0} \frac{\partial n}{\partial x} \int_0^{v_0} \frac{v^2\left(v_0-v\right)^2}{D} d v\,.
\end{equation}

\section{Spatial diffusion with non-zero lower plateau
velocity}\label{app:non-0_low_bound}

Let us consider the plateau in the electron distribution function and spectral energy density
including a constant lower bound $v_{\text{min}}$:
\begin{equation}\label{eq:plateau_vmin2}
  f^{0}\left(v,x,t\right) =\left\{
\begin{array}{ll}
p\left( x,t\right) , & v_{\text{min}}<v<u(x,t) \\
0, & v\leq v_{\text{min}}, \ v\geq u(x,t)%
\end{array}%
\right.
\end{equation}

\begin{equation}
W^{0}\left(v,x,t\right) =\left\{
\begin{array}{ll}
W_{0}\left(v,x,t\right) , &v_\text{min}<v<u(x,t) \\
0,& v\leq v_\text{min}, \ v\geq u(x,t)
\end{array}%
\right.
\end{equation}%

Solutions for $u(x,t)$, $p(x,t)$, and $W_0(x,v,t)$ can be found following
the method used in \citet{1998PlPhR..24..772K} to be

\begin{align}
&u(x,t)=v_0\,,\\
&p(x, t)=\frac{n_b}{v_0-v_{\text {min}}} \exp \left[-\frac{\left(x-\left(v_0+v_{\text {min}}\right) t / 2\right)^2}{d^2}\right]\,,\\
&W_{0}\left(v,x,t\right) =\frac{m}{\omega_{pe}} v^3\left(v-v_{\min }\right)\left(1-\frac{v+v_{\min }}{v_0+v_{\min }}\right) p(x, t)\,,
\end{align}
for the initial conditions on the electron distribution function given by
equation~(\ref{eq:ft0}).

Integrating equation~(\ref{eq:exp_eq}) over velocity from $v_{\min}$ to
$\infty$, one obtains
\begin{equation}\label{eq:f1_eq_vmin}
\frac{\partial n}{\partial t}+\frac{v_0+v_{\min}}{2}\frac{\partial n}{\partial x}+%
\frac{\partial }{\partial x}\int\limits_{v_{\min}}^{v_0}vf^{1}\text{d}v=0.
\end{equation}

Multiplying equation (\ref{eq:exp_eq}) by $v_0 - v_{\min}$ and subtracting
equation (\ref{eq:f1_eq_vmin}) one finds
\begin{equation}\label{eq:diff_deriv_vmin}
  \left( v-\frac{v_0+v_{\min}}{2}\right) \frac{\partial n}{\partial x}+(v_0-v_{\min})\frac{%
    \partial f^{1}}{\partial t}+(v_0-v_{\min})v\frac{\partial f^{1}}{\partial x}-\frac{%
    \partial }{\partial x}\int\limits_{v_{\min}}^{v_0}vf^{1}\text{d}v=(v_0-v_{\min})\frac{\partial }{%
    \partial v}D\frac{\partial f^{1}}{\partial v}.
\end{equation}%
and retaining only zero order terms
\begin{equation}\label{eq:diff_deriv2_vmin}
\frac{1}{v_0-v_{\min}}\left( v-\frac{v_0+v_{\min}}{2}\right) \frac{\partial n}{\partial x}=%
\frac{\partial }{\partial v}D\frac{\partial f^{1}}{\partial v}\,,
\end{equation}%
which is the equation for $f^1$.
Integrating equation (\ref{eq:diff_deriv2_vmin}) over $v$, one obtains
\begin{equation}
D\frac{\partial f^{1}}{\partial v}=\frac{1}{v_0-v_{\min}}\frac{\partial n}{\partial
x}\int\limits_{v_{\min}}^{v}\left( v^{^{\prime }}-\frac{v_0+v_{\min}}{2}\right)
\text{d}v^{^{\prime }}=\frac{1}{v_0-v_{\min}}\frac{\partial n}{\partial x}\frac{1}{2}%
\left(v - v_{\min}\right)\left( v-v_0\right) +C_{1}\,.
\end{equation}%
Since the quasilinear relaxation operates now on $v_{\min}<v<v_0$,
we have $\left. D\right\vert _{v=v_{\min}}=\left. D\right\vert _{v=v_0}=0$.
Therefore, $C_{1}=0$, yielding
\begin{equation}
\frac{\partial f^{1}}{\partial v}=\frac{(v-v_{\min})\left( v-v_0\right) }{2(v_0-v_{\min})D}%
\frac{\partial n}{\partial x}\,,
\end{equation}%
and with further integration over $v$, we find the expression for $f^1$
\begin{equation}\label{Append B}
f^{1}=\frac{1}{2(v_0-v_{\min})}\frac{\partial n}{\partial x}\int\limits_{v_{\min}}^{v}\frac{%
(v^{^{\prime}}-v_{\min})(v^{^{\prime}}-v_0)}{D}\text{d}v^{^{\prime }}+C_{2}\,,
\end{equation}
where the constant $C_2$ is determined from
\[
\int\limits_{v_{\min}}^{v_0}f^{1}\text{d}v=0= \left.\left[ v f^1\right]\right|^{v_0}_{v_{\min}} - \int^{v_0}\limits_{v_{\min}} v \frac{\partial f^1}{\partial v} \text{d}v\,\,,
\]
which can be expanded into
\[
\frac{1}{2(v_0-v_{\min})}\frac{\partial n}{\partial x}\int\limits_{v_{\min}}^{v_0} (v_0-v^{^{\prime}}) \frac{
(v^{^{\prime}}-v_{\min})(v^{^{\prime}}-v_0)}{D}\text{d}v^{^{\prime}}+(v_0-v_{\min})\,C_{2} = 0.
\]
Hence, the integration constant $C_2$ is found to be
\begin{equation}\label{eq:C2_vmin}
C_{2}=\frac{1}{2(v_0-v_{\min})^2}\frac{\partial n}{\partial x}
\int\limits_{v_{\min}}^{v_0} (v_0-v^{^{\prime}})^2
\frac{(v^{^{\prime}}-v_{\min})}{D}\text{d}v^{^{\prime}}\,,
\end{equation}
and $f^1$ can now be written as
\begin{equation}
f^{1}=\frac{1}{2(v_0-v_{\min})}\frac{\partial n}{\partial x}\left[ \int\limits_{v_{\min}}^{v}%
\frac{
(v^{^{\prime}}-v_{\min})(v^{^{\prime}}-v_0)}{D}\text{d}v^{^{\prime }}
+\frac{1}{v_0-v_{\min}}\int\limits_{v_{\min}}^{v_0}\left(v_0-v'\right)^2
\frac{(v^{^{\prime}}-v_{\min})}{D}\text{d}v'\right] .
\end{equation}
This allows us to rewrite the integral in the diffusion term of
equation~(\ref{eq:f1_eq_vmin}) as
\begin{equation}
\int\limits_{v_{\min}}^{v_0}vf^{1}\text{d}v=\left[ \frac{v^2}{2}f^1\right]^{v_0}_{v_{\min}} - \int^{v_0}\limits_{v_{\min}} \frac{v^2}{2} \frac{\partial f^1}{\partial v} \text{d}v\,\,
=-\frac{1}{4(v_0-v_{\text{min}})}\frac{\partial n}{\partial x}\int\limits_{v_{\text{min}}}^{v_0}%
\frac{(v-v_{\text{min}})^{2}\left( v_0-v\right) ^{2}}{D}\text{d}v.
\end{equation}

Using the diffusion coefficient in velocity space $D$ from
equation~(\ref{eq:D_v_vmin}), we have the spatial diffusion coefficient
\begin{align}
D_{xx} &= - \int^{v_0}\limits_{v_{\min}} v f^1 \text{d}v = \frac{1}{4(v_0 - v_{\min})} \int^{v_0}\limits_{v_{\min}} \frac{(v-v_{\min})^2 (v_0-v)^2}{D(v)} \text{d}v\\ \nonumber
&= \frac{v_0 + v_{\min}}{4D_0(v_0 - v_{\min})} \int^{v_0}\limits_{v_{\min}} \frac{(v-v_{\min}) (v_0-v)}{v^2} \text{d}v \\ \nonumber
&= \frac{v_0 + v_{\min}}{4D_0(v_0 - v_{\min})} \left.\left[ (v_0 + v_{\min}) \ln(v) + \frac{v_0 v_{\min}}{v} -v \right]\right\vert^{v_0}_{v_{\min}}\\ \nonumber
&=\frac{v_0 + v_{\min}}{4D_0}\left[ \left(\frac{v_0 + v_{\min}}{v_0 - v_{\min}}\right)\ln \left(\frac{v_0}{v_{\min}}\right) -2 \right]\,,
\end{align}
which is given in the main text by equation (\ref{eq:Dxx_vimin}).

\bibliographystyle{aasjournal}
\bibliography{refs_all}

\begin{thebibliography}{}
\expandafter\ifx\csname natexlab\endcsname\relax\def\natexlab#1{#1}\fi
\providecommand{\url}[1]{\href{#1}{#1}}
\providecommand{\dodoi}[1]{doi:~\href{http://doi.org/#1}{\nolinkurl{#1}}}
\providecommand{\doeprint}[1]{\href{http://ascl.net/#1}{\nolinkurl{http://ascl.net/#1}}}
\providecommand{\doarXiv}[1]{\href{https://arxiv.org/abs/#1}{\nolinkurl{https://arxiv.org/abs/#1}}}

\bibitem[{{Abramowitz} \& {Stegun}(1970)}]{1970hmfw.book.....A}
{Abramowitz}, M., \& {Stegun}, I.~A. 1970, {Handbook of mathematical functions
  : with formulas, graphs, and mathematical tables} (U.S. Dept. of Commerce,
  National Bureau of Standards)

\bibitem[{{Akbari} {et~al.}(2021){Akbari}, {LaBelle}, \&
  {Newman}}]{2021FrASS...7..116A}
{Akbari}, H., {LaBelle}, J.~W., \& {Newman}, D.~L. 2021, Frontiers in Astronomy
  and Space Sciences, 7, 116, \dodoi{10.3389/fspas.2020.617792}

\bibitem[{{Bardwell} \& {Goldman}(1976)}]{1976ApJ...209..912B}
{Bardwell}, S., \& {Goldman}, M.~V. 1976, \apj, 209, 912,
  \dodoi{10.1086/154790}

\bibitem[{{Barenblatt}(1996)}]{1996sssi.book.....B}
{Barenblatt}, G.~I. 1996, {Scaling, Self-similarity, and Intermediate
  Asymptotics} (Cambridge, UK: Cambridge University Press),
  \dodoi{10.1017/CBO9781107050242}

\bibitem[{{Benz}(2017)}]{2017LRSP...14....2B}
{Benz}, A.~O. 2017, Living Reviews in Solar Physics, 14, 2,
  \dodoi{10.1007/s41116-016-0004-3}

\bibitem[{Berryman \& Holland(1982)}]{Berryman82}
Berryman, J.~G., \& Holland, C.~J. 1982, Journal of Mathematical Physics, 23,
  983, \dodoi{10.1063/1.525466}

\bibitem[{{Chapman} \& {Cowling}(1970)}]{1970mtnu.book.....C}
{Chapman}, S., \& {Cowling}, T.~G. 1970, {The mathematical theory of
  non-uniform gases. an account of the kinetic theory of viscosity, thermal
  conduction and diffusion in gases} (Cambridge: University Press)

\bibitem[{{Che} {et~al.}(2017){Che}, {Goldstein}, {Diamond}, \&
  {Sagdeev}}]{2017PNAS..114.1502C}
{Che}, H., {Goldstein}, M.~L., {Diamond}, P.~H., \& {Sagdeev}, R.~Z. 2017,
  Proceedings of the National Academy of Science, 114, 1502,
  \dodoi{10.1073/pnas.1614055114}

\bibitem[{{Drummond} \& {Pines}(1964)}]{1964AnPhy..28..478D}
{Drummond}, W.~E., \& {Pines}, D. 1964, Annals of Physics, 28, 478,
  \dodoi{10.1016/0003-4916(64)90205-2}

\bibitem[{{Fainberg} \& {Stone}(1970)}]{1970SoPh...15..222F}
{Fainberg}, J., \& {Stone}, R.~G. 1970, \solphys, 15, 222,
  \dodoi{10.1007/BF00149487}

\bibitem[{Frank(2009)}]{Frank2009}
Frank, T.~D. 2009, Linear and Non-linear Fokker--Planck Equations (New York,
  NY: Springer New York), 5239--5265, \dodoi{10.1007/978-0-387-30440-3_311}

\bibitem[{{Frasca}(2008)}]{2008IJMPA..23..299F}
{Frasca}, M. 2008, International Journal of Modern Physics A, 23, 299,
  \dodoi{10.1142/S0217751X08038160}

\bibitem[{{Ginzburg} \& {Zhelezniakov}(1958)}]{1958SvA.....2..653G}
{Ginzburg}, V.~L., \& {Zhelezniakov}, V.~V. 1958, Soviet~Ast., 2, 653

\bibitem[{{Goldman} \& {Dubois}(1982)}]{1982PhFl...25.1062G}
{Goldman}, M.~V., \& {Dubois}, D.~F. 1982, Physics of Fluids, 25, 1062,
  \dodoi{10.1063/1.863839}

\bibitem[{{Grognard}(1982)}]{1982SoPh...81..173G}
{Grognard}, R.~J.~M. 1982, \solphys, 81, 173, \dodoi{10.1007/BF00151988}

\bibitem[{{Hannah} {et~al.}(2009){Hannah}, {Kontar}, \&
  {Sirenko}}]{2009ApJ...707L..45H}
{Hannah}, I.~G., {Kontar}, E.~P., \& {Sirenko}, O.~K. 2009, \apjl, 707, L45,
  \dodoi{10.1088/0004-637X/707/1/L45}

\bibitem[{{Hasselmann} \& {Wibberenz}(1970)}]{1970ApJ...162.1049H}
{Hasselmann}, K., \& {Wibberenz}, G. 1970, \apj, 162, 1049,
  \dodoi{10.1086/150736}

\bibitem[{Hill \& Hill(1993)}]{Hill93}
Hill, D.~L., \& Hill, J.~M. 1993, Quarterly of Applied Mathematics, 51, 633.
\newblock \url{http://www.jstor.org/stable/43637952}

\bibitem[{{Holman} {et~al.}(2011){Holman}, {Aschwanden}, {Aurass}, {Battaglia},
  {Grigis}, {Kontar}, {Liu}, {Saint-Hilaire}, \&
  {Zharkova}}]{2011SSRv..159..107H}
{Holman}, G.~D., {Aschwanden}, M.~J., {Aurass}, H., {et~al.} 2011, \ssr, 159,
  107, \dodoi{10.1007/s11214-010-9680-9}

\bibitem[{{Jeffrey} {et~al.}(2016){Jeffrey}, {Fletcher}, \&
  {Labrosse}}]{2016A&A...590A..99J}
{Jeffrey}, N. L.~S., {Fletcher}, L., \& {Labrosse}, N. 2016, \aap, 590, A99,
  \dodoi{10.1051/0004-6361/201527986}

\bibitem[{{Jokipii}(1966)}]{1966ApJ...146..480J}
{Jokipii}, J.~R. 1966, \apj, 146, 480, \dodoi{10.1086/148912}

\bibitem[{Juan R.~Esteban \&
  Vázquez(1988{\natexlab{a}})}]{doi:10.1080/03605308808820566}
Juan R.~Esteban, A.~R., \& Vázquez, J.~L. 1988{\natexlab{a}}, Communications
  in Partial Differential Equations, 13, 985, \dodoi{10.1080/03605308808820566}

\bibitem[{Juan R.~Esteban \& Vázquez(1988{\natexlab{b}})}]{Esteban88}
---. 1988{\natexlab{b}}, Communications in Partial Differential Equations, 13,
  985, \dodoi{10.1080/03605308808820566}

\bibitem[{{Kaplan} \& {Tsytovich}(1973)}]{1973plas.book.....K}
{Kaplan}, S.~A., \& {Tsytovich}, V.~N. 1973, {Plasma astrophysics} (Oxford:
  Pergamon Press)

\bibitem[{{Karlicky}(1997)}]{1997SSRv...81..143K}
{Karlicky}, M. 1997, \ssr, 81, 143, \dodoi{10.1023/A:1004939526282}

\bibitem[{Kheifets(1984)}]{Kheifets1982}
Kheifets, S. 1984, Part. Accel., 15, 67

\bibitem[{{King}(1993)}]{1993RSPTA.343..337K}
{King}, J.~R. 1993, Philosophical Transactions of the Royal Society of London
  Series A, 343, 337, \dodoi{10.1098/rsta.1993.0052}

\bibitem[{{Kontar}(2001{\natexlab{a}})}]{2001A&A...375..629K}
{Kontar}, E.~P. 2001{\natexlab{a}}, \aap, 375, 629,
  \dodoi{10.1051/0004-6361:20010807}

\bibitem[{{Kontar}(2001{\natexlab{b}})}]{2001CoPhC.138..222K}
---. 2001{\natexlab{b}}, Computer Physics Communications, 138, 222,
  \dodoi{10.1016/S0010-4655(01)00214-4}

\bibitem[{{Kontar} {et~al.}(1998){Kontar}, {Lapshin}, \&
  {Melnik}}]{1998PlPhR..24..772K}
{Kontar}, E.~P., {Lapshin}, V.~I., \& {Melnik}, V.~N. 1998, Plasma Physics
  Reports, 24, 772

\bibitem[{{Kontar} \& {P{\'e}cseli}(2002)}]{2002PhRvE..65f6408K}
{Kontar}, E.~P., \& {P{\'e}cseli}, H.~L. 2002, \pre, 65, 066408,
  \dodoi{10.1103/PhysRevE.65.066408}

\bibitem[{{Krafft} \& {Savoini}(2023)}]{2023ApJ...949...24K}
{Krafft}, C., \& {Savoini}, P. 2023, \apj, 949, 24,
  \dodoi{10.3847/1538-4357/acc1e4}

\bibitem[{{Krucker} {et~al.}(2007){Krucker}, {Kontar}, {Christe}, \&
  {Lin}}]{2007ApJ...663L.109K}
{Krucker}, S., {Kontar}, E.~P., {Christe}, S., \& {Lin}, R.~P. 2007, \apjl,
  663, L109, \dodoi{10.1086/519373}

\bibitem[{{Krupar} {et~al.}(2014){Krupar}, {Maksimovic}, {Santolik}, {Kontar},
  {Cecconi}, {Hoang}, {Kruparova}, {Soucek}, {Reid}, \&
  {Zaslavsky}}]{2014SoPh..289.3121K}
{Krupar}, V., {Maksimovic}, M., {Santolik}, O., {et~al.} 2014, \solphys, 289,
  3121, \dodoi{10.1007/s11207-014-0522-x}

\bibitem[{{Li} {et~al.}(2008){Li}, {Cairns}, \&
  {Robinson}}]{2008JGRA..113.6104L}
{Li}, B., {Cairns}, I.~H., \& {Robinson}, P.~A. 2008, Journal of Geophysical
  Research (Space Physics), 113, A06104, \dodoi{10.1029/2007JA012957}

\bibitem[{{Lin}(1970)}]{1970SoPh...12..266L}
{Lin}, R.~P. 1970, \solphys, 12, 266, \dodoi{10.1007/BF00227122}

\bibitem[{{Lin}(1974)}]{1974SSRv...16..189L}
---. 1974, \ssr, 16, 189, \dodoi{10.1007/BF00240886}

\bibitem[{{Lin}(1985)}]{1985SoPh..100..537L}
---. 1985, \solphys, 100, 537, \dodoi{10.1007/BF00158444}

\bibitem[{{Lonngren} \& {Hirose}(1976)}]{1976PhLA...59..285L}
{Lonngren}, K.~E., \& {Hirose}, A. 1976, Physics Letters A, 59, 285,
  \dodoi{10.1016/0375-9601(76)90794-5}

\bibitem[{{Lyubchyk} {et~al.}(2017){Lyubchyk}, {Kontar}, {Voitenko}, {Bian}, \&
  {Melrose}}]{2017SoPh..292..117L}
{Lyubchyk}, O., {Kontar}, E.~P., {Voitenko}, Y.~M., {Bian}, N.~H., \&
  {Melrose}, D.~B. 2017, \solphys, 292, 117, \dodoi{10.1007/s11207-017-1140-1}

\bibitem[{{Magelssen} \& {Smith}(1977)}]{1977SoPh...55..211M}
{Magelssen}, G.~R., \& {Smith}, D.~F. 1977, \solphys, 55, 211,
  \dodoi{10.1007/BF00150886}

\bibitem[{{Mel'nik}(1995)}]{1995PlPhR..21...89M}
{Mel'nik}, V.~N. 1995, Plasma Physics Reports, 21, 89,
  \dodoi{10.48550/arXiv.1802.07806}

\bibitem[{{Mel'nik} \& {Kontar}(2000)}]{2000NewA....5...35M}
{Mel'nik}, V.~N., \& {Kontar}, E.~P. 2000, \na, 5, 35,
  \dodoi{10.1016/S1384-1076(00)00004-X}

\bibitem[{{Mel'nik} {et~al.}(1999){Mel'nik}, {Lapshin}, \&
  {Kontar}}]{1999SoPh..184..353M}
{Mel'nik}, V.~N., {Lapshin}, V., \& {Kontar}, E. 1999, \solphys, 184, 353,
  \dodoi{10.1023/A:1005191910544}

\bibitem[{{Muschietti}(1990)}]{1990SoPh..130..201M}
{Muschietti}, L. 1990, \solphys, 130, 201, \dodoi{10.1007/BF00156790}

\bibitem[{{Muschietti} {et~al.}(1985){Muschietti}, {Goldman}, \&
  {Newman}}]{1985SoPh...96..181M}
{Muschietti}, L., {Goldman}, M.~V., \& {Newman}, D. 1985, \solphys, 96, 181,
  \dodoi{10.1007/BF00239800}

\bibitem[{{Okubo} {et~al.}(1984){Okubo}, {Mitchell}, \&
  {Andreasen}}]{1984PhLA..105..169O}
{Okubo}, A., {Mitchell}, J., \& {Andreasen}, V. 1984, Physics Letters A, 105,
  169, \dodoi{10.1016/0375-9601(84)90389-X}

\bibitem[{{Papadopoulos} {et~al.}(1974){Papadopoulos}, {Goldstein}, \&
  {Smith}}]{1974ApJ...190..175P}
{Papadopoulos}, K., {Goldstein}, M.~L., \& {Smith}, R.~A. 1974, \apj, 190, 175,
  \dodoi{10.1086/152862}

\bibitem[{{Pedron} {et~al.}(2005){Pedron}, {Mendes}, {Buratta}, {Malacarne}, \&
  {Lenzi}}]{2005PhRvE..72c1106P}
{Pedron}, I.~T., {Mendes}, R.~S., {Buratta}, T.~J., {Malacarne}, L.~C., \&
  {Lenzi}, E.~K. 2005, \pre, 72, 031106, \dodoi{10.1103/PhysRevE.72.031106}

\bibitem[{{Posner}(2007)}]{2007SpWea...5.5001P}
{Posner}, A. 2007, Space Weather, 5, 05001, \dodoi{10.1029/2006SW000268}

\bibitem[{{Ratcliffe} {et~al.}(2014){Ratcliffe}, {Kontar}, \&
  {Reid}}]{2014A&A...572A.111R}
{Ratcliffe}, H., {Kontar}, E.~P., \& {Reid}, H.~A.~S. 2014, \aap, 572, A111,
  \dodoi{10.1051/0004-6361/201423731}

\bibitem[{{Reid} \& {Kontar}(2013)}]{2013SoPh..285..217R}
{Reid}, H. A.~S., \& {Kontar}, E.~P. 2013, \solphys, 285, 217,
  \dodoi{10.1007/s11207-012-0013-x}

\bibitem[{{Reid} \& {Kontar}(2018)}]{2018A&A...614A..69R}
---. 2018, \aap, 614, A69, \dodoi{10.1051/0004-6361/201732298}

\bibitem[{{Reid} {et~al.}(2014){Reid}, {Vilmer}, \&
  {Kontar}}]{2014A&A...567A..85R}
{Reid}, H. A.~S., {Vilmer}, N., \& {Kontar}, E.~P. 2014, \aap, 567, A85,
  \dodoi{10.1051/0004-6361/201321973}

\bibitem[{{Rosenau}(1995)}]{1995PhRvL..74.1056R}
{Rosenau}, P. 1995, \prl, 74, 1056, \dodoi{10.1103/PhysRevLett.74.1056}

\bibitem[{{Ryutov}(2018)}]{2018PhPl...25j0501R}
{Ryutov}, D.~D. 2018, Physics of Plasmas, 25, 100501, \dodoi{10.1063/1.5042254}

\bibitem[{{Ryutov} \& {Sagdeev}(1970)}]{1970JETP...31..396R}
{Ryutov}, D.~D., \& {Sagdeev}, R.~Z. 1970, Soviet Journal of Experimental and
  Theoretical Physics, 31, 396

\bibitem[{{Sauer} {et~al.}(2019){Sauer}, {Baumg{\"a}rtel}, {Sydora}, \&
  {Winterhalter}}]{2019JGRA..124...68S}
{Sauer}, K., {Baumg{\"a}rtel}, K., {Sydora}, R., \& {Winterhalter}, D. 2019,
  Journal of Geophysical Research (Space Physics), 124, 68,
  \dodoi{10.1029/2018JA025887}

\bibitem[{{Schlickeiser}(1989)}]{1989ApJ...336..243S}
{Schlickeiser}, R. 1989, \apj, 336, 243, \dodoi{10.1086/167009}

\bibitem[{{Sturrock}(1964)}]{1964NASSP..50..357S}
{Sturrock}, P.~A. 1964, NASA Special Publication, 50, 357

\bibitem[{{Takakura}(1982)}]{1982SoPh...78..141T}
{Takakura}, T. 1982, \solphys, 78, 141, \dodoi{10.1007/BF00151150}

\bibitem[{{Takakura} \& {Shibahashi}(1976)}]{1976SoPh...46..323T}
{Takakura}, T., \& {Shibahashi}, H. 1976, \solphys, 46, 323,
  \dodoi{10.1007/BF00149860}

\bibitem[{{Timofeev} {et~al.}(2015){Timofeev}, {Annenkov}, \&
  {Arzhannikov}}]{2015PhPl...22k3109T}
{Timofeev}, I.~V., {Annenkov}, V.~V., \& {Arzhannikov}, A.~V. 2015, Physics of
  Plasmas, 22, 113109, \dodoi{10.1063/1.4935890}

\bibitem[{{Treumann}(1997)}]{1997GeoRL..24.1727T}
{Treumann}, R.~A. 1997, \grl, 24, 1727, \dodoi{10.1029/97GL01760}

\bibitem[{V{\'a}zquez(2017)}]{2017arXiv170608241V}
V{\'a}zquez, J.~L. 2017, The Mathematical Theories of Diffusion: Nonlinear and
  Fractional Diffusion (Cham: Springer International Publishing), 205--278,
  \dodoi{10.1007/978-3-319-61494-6_5}

\bibitem[{{Vedenov} {et~al.}(1967){Vedenov}, {Gordeev}, \&
  {Rudakov}}]{1967PlPh....9..719V}
{Vedenov}, A.~A., {Gordeev}, A.~V., \& {Rudakov}, L.~I. 1967, Plasma Physics,
  9, 719, \dodoi{10.1088/0032-1028/9/6/305}

\bibitem[{{Vedenov} \& {Velikhov}(1963)}]{1963JETP...16..682V}
{Vedenov}, A.~A., \& {Velikhov}, E.~P. 1963, Soviet Journal of Experimental and
  Theoretical Physics, 16, 682

\bibitem[{{Vedenov} {et~al.}(1961){Vedenov}, {Velikhov}, \&
  {Sagdeev}}]{1961SvPhU...4..332V}
{Vedenov}, A.~A., {Velikhov}, E.~P., \& {Sagdeev}, R.~Z. 1961, Soviet Physics
  Uspekhi, 4, 332, \dodoi{10.1070/PU1961v004n02ABEH003341}

\bibitem[{Whiting(1968)}]{WHITING19681379}
Whiting, E. 1968, Journal of Quantitative Spectroscopy and Radiative Transfer,
  8, 1379, \dodoi{https://doi.org/10.1016/0022-4073(68)90081-2}

\bibitem[{{Yoon} {et~al.}(2012){Yoon}, {Ziebell}, {Gaelzer}, {Lin}, \&
  {Wang}}]{2012SSRv..173..459Y}
{Yoon}, P.~H., {Ziebell}, L.~F., {Gaelzer}, R., {Lin}, R.~P., \& {Wang}, L.
  2012, \ssr, 173, 459, \dodoi{10.1007/s11214-012-9867-3}

\bibitem[{{Zaitsev} {et~al.}(1972){Zaitsev}, {Mityakov}, \&
  {Rapoport}}]{1972SoPh...24..444Z}
{Zaitsev}, V.~V., {Mityakov}, N.~A., \& {Rapoport}, V.~O. 1972, \solphys, 24,
  444, \dodoi{10.1007/BF00153387}

\bibitem[{{Zheleznyakov} \& {Zaitsev}(1970)}]{1970SvA....14...47Z}
{Zheleznyakov}, V.~V., \& {Zaitsev}, V.~V. 1970, \sovast, 14, 47

\bibitem[{{Ziebell} {et~al.}(2008){Ziebell}, {Gaelzer}, \&
  {Yoon}}]{2008PhPl...15c2303Z}
{Ziebell}, L.~F., {Gaelzer}, R., \& {Yoon}, P.~H. 2008, Physics of Plasmas, 15,
  032303, \dodoi{10.1063/1.2844740}

\bibitem[{{Ziebell} {et~al.}(2011){Ziebell}, {Yoon}, {Pavan}, \&
  {Gaelzer}}]{2011PPCF...53h5004Z}
{Ziebell}, L.~F., {Yoon}, P.~H., {Pavan}, J., \& {Gaelzer}, R. 2011, Plasma
  Physics and Controlled Fusion, 53, 085004,
  \dodoi{10.1088/0741-3335/53/8/085004}

\bibitem[{{Zimbardo} {et~al.}(2006){Zimbardo}, {Pommois}, \&
  {Veltri}}]{2006ApJ...639L..91Z}
{Zimbardo}, G., {Pommois}, P., \& {Veltri}, P. 2006, \apjl, 639, L91,
  \dodoi{10.1086/502676}

\end{thebibliography}

\end{document}